\documentclass[letterpaper,twocolumn,showpacs,pra,aps]{revtex4-1}

\usepackage{amsmath}  
\usepackage{amsfonts} 
\usepackage{graphicx} 
\usepackage{amssymb}
\usepackage{url}

\begin{document}

\title{Tunneling Newtonian Universe}

\author{Eugene B. Kolomeisky}

\affiliation
{Department of Physics, University of Virginia, P. O. Box 400714, Charlottesville, Virginia 22904-4714, USA}

\date{\today}

\begin{abstract}
We analyze quantum-mechanical counterpart of Newtonian cosmology and show that effects of zero-point motion eliminate classical density singularity.  Quantum effects are particularly significant for closed Universes where without the cosmological constant the energy spectrum and space curvature are quantized.  When small positive cosmological constant is included, these states become quasi-stationary and decay via tunneling.  Corresponding metastable Universes evolve in three stages:  long period of gestation followed by rapid tunneling expansion further followed by slower Hubble expansion.  This closely resembles inflation scenario of modern cosmology.

\end{abstract}


\maketitle

\section{Introduction}

Following discovery of an expansion of the Universe \cite{Hubble, Friedmann,Lem}, it was realized that naive extrapolation back in time of the equations of the theory leads to a density singularity called the Big Bang.   It is believed that the singularity is an artifact of applying the general theory of relativity beyond its range of applicability.   It would be avoided in a quantum theory of gravity, arguably the biggest unsolved problem of fundamental physics.  

The early Universe is not the only setting where the general theory of relativity and quantum mechanics clash - singularities, unavoidable in Einstein's theory \cite{Penrose}, are also found in the interior of black holes.  The need for a quantum theory of gravity is not limited to taming of singularities \cite{Kefir}.  One of the obstacles to progress is a lack of empirical input.  Indeed, quantum effects should play a role on the Planck scales of length, $l_{P}=\sqrt{\hbar G/c^{3}}\approx 1.62\times10^{-33}$ cm, and time, $t_{P}=\sqrt{\hbar G/c^{5}}\approx5.39\times10^{-44}$ s, which are unlikely to be experimentally probed.  

Therefore understanding the early Universe involves conjectures about new physics that are difficult to scrutinize as targeted experimentation is impossible.  Perhaps that is why the dominant hypothesis, the inflation, a short period of accelerated expansion of the early Universe due to decay of a scalar field called the inflaton \cite{Harrison,Mukhanov}, is opposed by a number of researchers \cite{Turok,opposition}.

In such circumstances, making a progress even on a simplest, level, i.e. combining non-relativistic quantum mechanics with Newtonian cosmology in a logically consistent manner may prove useful.  While what is described below must be viewed as a toy model, qualitatively our predictions resemble the inflation scenario if positive cosmological constant, the driving force of the evolution, is \textit{small} \cite{Harrison,Mukhanov}.  The physical picture we arrive at represents an illustration of Coleman's idea of decay of a "false vacuum" \cite{Coleman}.  Moreover, we believe that in the limit when the Einstein's theory of general relativity reduces to Newtonian gravity \cite{LL2}, a future consistent theory of quantum cosmology would reduce to ours.

The classical theory whose quantum counterpart is analyzed below is due to Milne and McCrea  \cite{Milne,MM,McCrea0,McCrea1} who demonstrated that non-relativistic version of Friedmann's cosmological equations can be directly obtained from Newtonian gravity.  There have been several attempts in the past to analyze quantum version of this theory \cite{Zamora,Vieira15,Vieira19,Gouba}.  Our treatment while relying on previous results \cite{Milne,MM,McCrea0,McCrea1,Zamora,Vieira15, Vieira19,Gouba,EBK,Bonnor}, focuses on questions pertinent to the early Universe, specifically, initial density singularity and inflation. 

\section{Review of Newtonian Cosmology}

We begin with a survey of Newtonian cosmology following Refs.\cite{Milne,MM,McCrea0,McCrea1,Vieira15,Vieira19,EBK,Bonnor}.  Let us consider an ideal non-relativistic gravitating liquid described by the position- and time-dependent number density $n(\textbf{r},t)$ and velocity $\textbf{v}(\textbf{r},t)$ fields. They are related by the continuity equation
\begin{equation}
\label{continuity}
\frac{\partial n}{\partial t}+\nabla \cdot(n\textbf{v})=0.
\end{equation}
The motion of the liquid is governed by the Euler equation 
\begin{equation}
\label{gravitational_Euler}
\frac{\partial \textbf{v}}{\partial t} +(\textbf{v}\cdot \nabla)\textbf{v}= \mathbf{g}-\frac{1}{mn}\nabla p
\end{equation} 
where $m$ is the particle mass, $p$ is the pressure, and the gravitational field $\mathbf{g}$ obeys Gauss's law \cite{MM,Bonnor}
\begin{equation}
\label{gravitational_Gauss}
\nabla\cdot\textbf{g}=-4\pi Gm(n-n_{0}),~~~n_{0}=\frac{\Lambda c^{2}}{4\pi G m}
\end{equation}
where the characteristic number density $n_{0}$ is due to the presence of Einstein's cosmological constant $\Lambda$;  Einstein's static model of the Universe is the $n=n_{0}$ case \cite{MM,Bonnor}.  

We seek solutions to Eqs.(\ref{continuity})-(\ref{gravitational_Gauss}) satisfying the cosmological principle, i.e. corresponding to a spatially homogeneous and isotropic matter density $n=n(t)$.  Then at any instant $t$ the pressure $p$ is also uniform, so that $\nabla p=0$.  The space isotropy dictates that in the rest frame of one of the particles the velocity field is radially symmetric, $\textbf{v}(\textbf{r},t)=v(r,t)\textbf{r}/r$, where the radius vector $\textbf{r}$ is the position relative to the particle at rest.    Then the dynamics of the liquid is governed by the gravitational field following from Gauss's law (\ref{gravitational_Gauss}):
\begin{equation}
\label{electric_field}
\textbf{g}=-\frac{4\pi Gm}{3}[n(t)-n_{0}]\textbf{r}.
\end{equation}
With all this in mind, the continuity (\ref{continuity}) and the Euler (\ref{gravitational_Euler}) equations can be written as
\begin{equation}
\label{radial_continuity}
\frac{\dot{n}}{n}+\frac{(r^{2}v)'}{r^{2}}=0
\end{equation} 
\begin{equation}
\label{radial_Euler}
\frac{\dot{v}+vv'}{r}=-\frac{4\pi Gm}{3}(n-n_{0})
\end{equation}
where the dot and the accent are shorthands for the derivatives with respect to $t$ and $r$.  Introducing a new function $H(t)$ according to 
\begin{equation}
\label{Hubble_definition}
\dot{n}=-3H(t)n
\end{equation}
allows one to integrate the continuity equation (\ref{radial_continuity}) with the result
\begin{equation}
\label{Hubble_law}
\textbf{v}=H(t)\textbf{r}
\end{equation}
known as the Hubble law;  $H(t)$ is the Hubble parameter. 

Substituting Eq.(\ref{Hubble_law}) into the Euler equation (\ref{radial_Euler}) we find 
\begin{equation}
\label{Friedmann2}
\dot{H}+H^{2}(t)=-\frac{4\pi Gm}{3}(n-n_{0})
\end{equation}
which is known as the Newton equation for the Universe.

While the evolution described by Eqs.(\ref{Hubble_definition})-(\ref{Friedmann2}) refers to the rest reference frame of one of the particles, one can show that the motion looks the same no matter where the origin is chosen \cite{ZN,McCrea1}.  Eqs.(\ref{Hubble_definition})-(\ref{Friedmann2}) capture the Friedmann cosmology in the non-relativistic regime \cite{MM,ZN}.  

Further progress can be made by introduction of a new function $a(t)$ called the scale factor \cite{ZN} such as 
\begin{equation}
\label{scale_factor}
H(t)=\frac{\dot{a}}{a}.
\end{equation}
In the Friedmann's cosmology $a(t)$ describes expansion or contraction of space itself.   

Combining Eqs.(\ref{scale_factor}) and (\ref{Hubble_definition}), the latter can be integrated with the result 
\begin{equation}
\label{charge_conservation}
n(t)=\frac{3N}{4\pi a^{3}(t)}
\end{equation}
where $N$ is a conserved number of particles within evolving sphere of radius $a(t)$.

\begin{figure}
\begin{center}
\includegraphics[width=1\columnwidth]{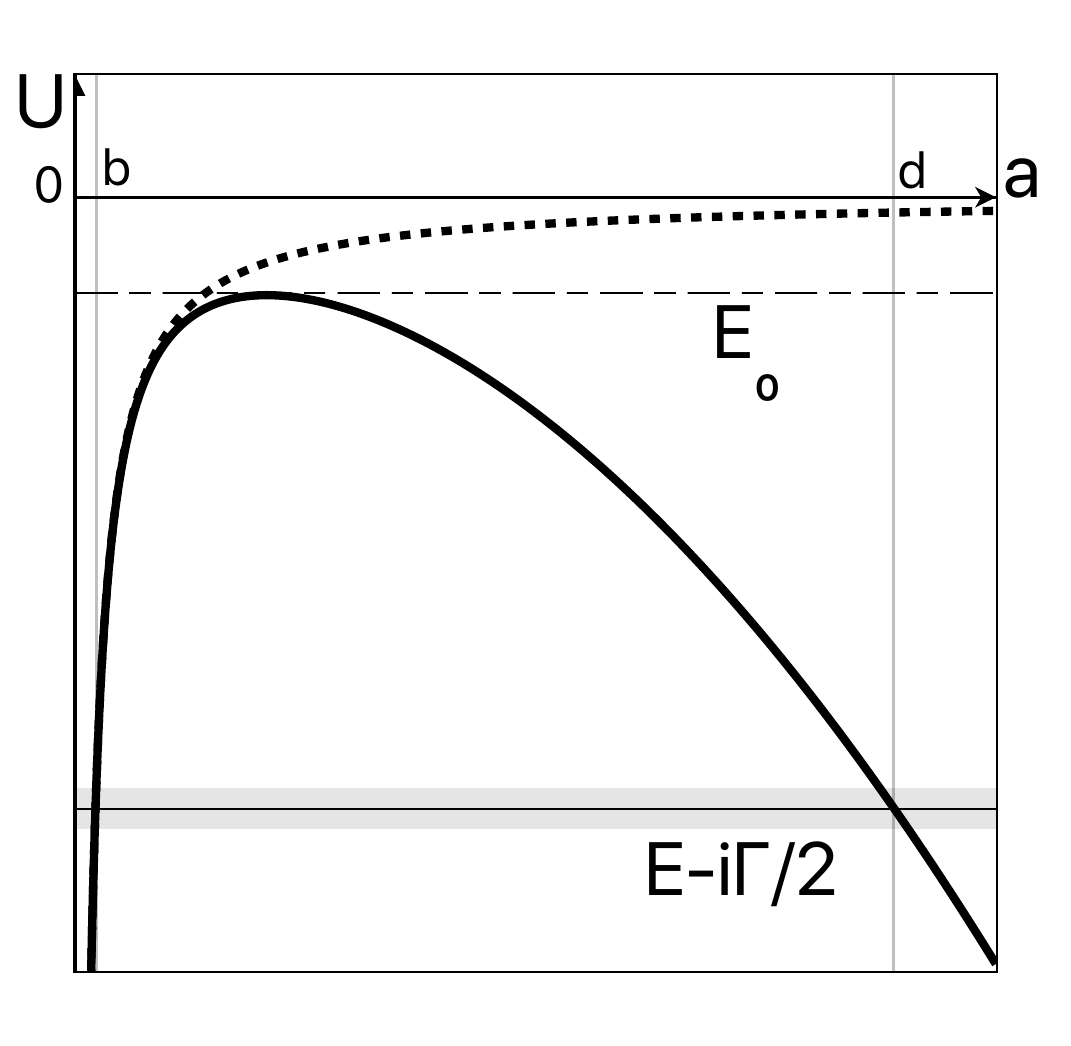}
\caption{Sketches of the potential energy $U(a)$,  Eqs. (\ref{potential_energy0}), (\ref{potential_energy1}) without the cosmological constant (dashed), and with it (solid).  The turning points of the classical motion $a=b$ and $a=d$ are solutions to the equation $U(a)=E$. Opaque band around the $E=U$ line symbolizes width $\Gamma$ of a quasi-stationary state in the presence of a small cosmological constant.  The line $U=E_{0}$ corresponds to Einstein's static model of the Universe.}
\label{pe}
\end{center}
\end{figure}

Substituting Eqs.(\ref{scale_factor}) and (\ref{charge_conservation}) into the Newton equation for the Universe (\ref{Friedmann2}), we find an expression
\begin{eqnarray}
\label{scale_factor_equation0}
m\ddot{a}&=&-\frac{4\pi Gm^{2}(n-n_{0})a^{3}}{3a^{2}}\nonumber\\
&=&-\frac{Gm^{2}N}{a^{2}}+\frac{4\pi}{3} Gm^{2}n_{0}a\equiv F(a)
\end{eqnarray}
that is the equation of motion for a zero angular momentum particle on the surface of an expanding or contracting sphere of radius $a$ caused by the force $F(a)$:  the particle is attracted by all the particles inside the sphere and repelled by the background due to the cosmological constant.  The energy integral of Eq.(\ref{scale_factor_equation0}) has the form 
\begin{equation}
\label{energy_integral0}
\frac{m\dot{a}^{2}}{2}+U(a)=E,
\end{equation}
\begin{equation}
\label{potential_energy0}
U(a)=-\int F(a)da=-\frac{Gm^{2}N}{a}-\frac{2\pi}{3}Gm^{2}n_{0}a^{2}
\end{equation}
where $U(a)$ is the potential energy and $E$ is the energy (Figure \ref{pe}).  The conservation laws (\ref{charge_conservation}), (\ref{energy_integral0}) and (\ref{potential_energy0}) can be combined into an expression \cite{ZN,MM,McCrea0,Bonnor}
\begin{equation}
\label{1stFriedmann0}
\left (\frac{\dot{a}}{a}\right )^{2}=\frac{8\pi Gm}{3}\left (n+\frac{n_{0}}{2}\right )-\frac{kc^{2}}{a^{2}}, ~~~k=-\frac{2E}{mc^{2}}
\end{equation}  
In the Friedmann's cosmology the combination $k/a^{2}$ in the last term in (\ref{1stFriedmann0}) is called the Gaussian curvature of space:  the $k>0$ space possesses positive curvature and is closed, the $k<0$ space is negatively curved and open, while $k=0$ describes Euclidian (flat) space. 
     
Classical dynamics of the scale factor $a(t)$ accumulated in Eq. (\ref{energy_integral0}) can be understood in terms of the central field motion of a particle of mass $m$, position $a$, and energy $E$ in the field of the potential energy $U(a)$ (\ref{potential_energy0}).  For any energy $E$ there are solutions to Eqs.(\ref{energy_integral0}) and (\ref{potential_energy0}) reaching $a=0$ which, in view of the number conservation law (\ref{charge_conservation}), imply density singularities.  

\section{Quantized Newtoninan Cosmology}

Since the classical dynamics of the scale factor $a(t)$ corresponds to the zero angular momentum Coulomb problem, we assert that the quantum-mechanical Hamiltonian of the problem is that of the central-field motion
\begin{equation}
\label{Hamiltonian}
\hat{\mathcal{H}}=-\frac{\hbar^{2}}{2m}\frac{1}{a^{2}}\frac{d}{da}\left (a^{2}\frac{d}{da}\right )+U(a)
\end{equation}
with only the radial part of the Laplacian in spherical coordinates included \cite{LL3}.  The Schr\"odinger equation has the form
\begin{equation}
\label{SE}
i\hbar\frac{\partial\Psi}{\partial t}=\hat{\mathcal{H}}\Psi
\end{equation}
where the wave function $\Psi(a,t)$ is the probability amplitude to have the scale factor $a$.  Since the potential energy $U(a)$ (\ref{potential_energy0}) is time-independent, one has
\begin{equation}
\label{wave_function}
\Psi(a,t)=\exp\left (-\frac{i}{\hbar}Et\right ) \psi(a)
\end{equation}
and the stationary-state wave function $\psi(a)$ is a solution to the Schr\"odinger equation
\begin{equation}
\label{SE1}
\frac{1}{a^{2}}\frac{d}{da}\left (a^{2}\frac{d\psi}{da}\right)+\frac{2m}{\hbar^{2}}[E-U(a)]\psi=0.
\end{equation}
It is a well-known property of the central-field motion that via the substitution 
\begin{equation}
\label{substitution}
\psi(a)=\frac{\chi(a)}{a}
\end{equation} 
Eq.(\ref{SE1}) reduces to the one-dimensional Schr\"odinger equation \cite{LL3}
\begin{equation}
\label{SE2}
\frac{d^{2}\chi}{da^{2}}+\frac{2m}{\hbar^{2}}[E-U(a)]\chi=0
\end{equation}
supplemented by the boundary condition $\chi(0)=0$.  The probability to have the scale factor in the range between $a$ and $a+da$ is given by $a^{2}|\psi(a)|^{2}da=|\chi(a)|^{2}da$.

The effect of zero-point motion can be understood by employing the uncertainty principle:  momentum conjugated to the scale factor $a$ cannot  be smaller than about $\hbar/a$.  This contributes kinetic energy of order $\hbar^{2}/ma^{2}$ to the potential energy (\ref{potential_energy0}).  The outcome may be viewed as an effective potential energy estimated as
\begin{equation}
\label{effective_potential_energy}
U_{eff}(a)\simeq \frac{\hbar^{2}}{ma^{2}}+ U(a).
\end{equation}
The latter, as $a\rightarrow 0$, is dominated by the zero-point energy behaving as  $1/a^{2}$, and the equation $E=U_{eff}(a)$ no longer has $a=0$ solutions \cite{LL3}.  Therefore classical density singularity is removed by the effects of zero-point motion.  

\subsection{Newtonian units}

As in the analysis of the motion in a Coulomb field \cite{LL3}, it will be convenient to use special units for the measurement of all quantities, that we call \textit{Newtonian units}.  As units of mass, length and time we take respectively,
\begin{eqnarray}
\label{primary_units}
m, a_{\mathcal{N}}&=&\frac{\hbar^{2}}{Gm^{3}N}=\frac{\hbar}{mc}\frac{N_{\mathcal{G}}}{N},\nonumber\\
t_{\mathcal{N}}&=&\frac{\hbar^{3}}{G^{2}m^{5}N^{2}}=\frac{\hbar}{mc^{2}}\left (\frac{N_{\mathcal{G}}}{N}\right )^{2}
\end{eqnarray}
where the parameter
\begin{equation}
\label{fine_structure}
N_{\mathcal{G}}=\frac{\hbar c}{Gm^{2}}\simeq10^{38}
\end{equation}
is the inverse of the gravitational fine structure constant.  To be definite hereafter we assume that gravitating liquid is made of particles of a nucleon mass, $m\simeq 10^{-24}$ g.  

The two representations for $a_{\mathcal{N}}$ in Eq.(\ref{primary_units}), the size of the ground-state wave function of the Newtonian Universe, feature two vastly different length scales:  $\hbar^{2}/Gm^{3}\simeq 10^{25}$ cm, the gravitational Bohr radius, that is only four orders of magnitude smaller than the size of the observable Universe \cite{Harrison,CMB}, and $\hbar/mc\simeq 10^{-13}$ cm, nucleon's Compton wavelength.  The second representation for the unit of time $t_{\mathcal{N}}$ features a time scale $\hbar/mc^{2}\simeq 10^{-23}$ s which is a time it takes light to travel a distance of nucleon's Compton wave length.

Given primary units (\ref{primary_units}), the remaining units can be derived accordingly.  Specifically, the units of velocity and energy will be given by 
\begin{equation}
\label{velocity_energy_units}
v_{\mathcal{N}}=\frac{a_{\mathcal{N}}}{t_{\mathcal{N}}}=c\frac{N}{N_{\mathcal{G}}},~~~~~ E_{\mathcal{N}}=\frac{\hbar^{2}}{ma_{N}^{2}}=mc^{2}\left (\frac{N}{N_{\mathcal{G}}}\right )^{2}.
\end{equation}

In these new units the potential energy (\ref{potential_energy0}) (sketched in Figure \ref{pe}) acquires the form
\begin{equation}
\label{potential_energy1}
U(a)=-\frac{1}{a}-\frac{\Omega^{2}a^{2}}{2},~\Omega=\sqrt{\frac{\Lambda}{3}} \frac{\hbar}{mc}\left (\frac{N_{\mathcal{G}}}{N}\right )^{2}.
\end{equation}

\subsection{Stationary states}

If $\Omega=0$, we have a well-known problem of motion of a zero angular momentum particle in a Coulomb field \cite{LL3}:  the wave functions are documented, the spectrum is continuous if $E>0$ and discrete if $E<0$.  In the latter case it is just the Bohr spectrum \cite{Zamora}
\begin{equation}
\label{Bohr}
E_{n_{r}}=-\frac{1}{2(n_{r}+1)^{2}}, ~n_{r}=0,1,2,...
\end{equation}    
where $n_{r}$ is the radial quantum number.  Since the energy $E$ determines the sign of the curvature coefficient entering Eq.(\ref{1stFriedmann0}), Eq.(\ref{Bohr}) implies quantization of the curvature coefficient, $k_{\nu}\propto1/(n_{r}+1)^{2}$, for closed Universes.  

Requiring that magnitudes of all the energies (expressed in the original physical units) are significantly smaller than the rest energy $mc^{2}$ or, equivalently, that all the velocities are much smaller than the speed of light (see Eqs.(\ref{velocity_energy_units}) and (\ref{Bohr})) leads to the constraint $N\ll N_{\mathcal{G}}$ which is the range of applicability of our theory.  It is important to keep in mind that the assumption of small velocity also implies that the gravitational field itself is weak, i.e. applicability of the Newtonian gravity \cite{LL2}. The condition $N\ll N_{\mathcal{G}}$ is violated for the physical Universe as the number of nucleons in the observable Universe, $N\simeq 10^{80}$ \cite{Harrison,CMB}, significantly exceeds $N_{\mathcal{G}}\simeq 10^{38}$ (\ref{fine_structure}).  That is why quantum Newtonian cosmology has a toy character and its predictions do not literally apply to the physical Universe.  

\subsection{Quasi-stationary states and tunneling}

For $\Omega$ finite the potential energy (\ref{potential_energy1}) has a maximum at $a_{0}=\Omega^{-2/3}$ and $E_{0}=-(3/2)\Omega^{2/3}$ (Figure \ref{pe}).  While classically the $E>E_{0}$ motion is infinite, quantum-mechanically there is a probability of over the barrier reflection.  Likewise, classically for $E<E_{0}$ there are two types of motion - finite for $0\leqslant a \leqslant b$ and infinite for $a\geqslant d$ where $b$ and $d$, solutions to the equation $U(a)=E$, are classical turning points.  Quantum-mechanically these two types of motion are connected by tunneling:  the system originally found at $0\leqslant a\leqslant b$ can tunnel  into the $a\geqslant d$ region.  Whatever the values of $E$ and $\Omega$, the spectrum is continuous, and the motion is semiclassical for $a$ large \cite{LL3}.  The classical momentum is then 
\begin{equation}
\label{momentum}
p(a)=\sqrt{2[E-U(a)]}\approx \Omega a+\frac{E}{\Omega a},
\end{equation}    
and the asymptotic form of solutions of the Schr\"odinger equations (\ref{SE1}) and (\ref{SE2}) is
\begin{eqnarray}
\label{asymptotic_solutions}
\psi(a)&=&\frac{\chi(a)}{a}\sim \frac{1}{a\sqrt{p(a)}}\exp\left (\pm i\int^{a} p(a')da'\right )\nonumber\\
&\sim& a^{\pm iE/\Omega-3/2}\exp\left (\pm i\frac{\Omega a^{2}}{2}\right )
\end{eqnarray}

Using the observed value of the cosmological constant $\Lambda = (1.106 \pm 0.023) \times 10^{-56}$ $\text{cm}^{-2}$ \cite{CMB} and assuming $N\simeq N_{\mathcal{G}}$, the dimensionless parameter $\Omega$ entering Eq.(\ref{potential_energy1}) can be estimated as $\Omega\simeq 10^{-41}\ll 1$.  The fact that it is much smaller than unity has a profound effect on the dynamics of the scale factor when $E\ll E_{0}$.  Now the turning points of the classical motion are approximately $b=1/|E|$ and $d=(2|E|)^{1/2}/\Omega\gg b$, i.e. the barrier is wide, the tunneling probability is small, and the concept of quasi-stationary states applies \cite{LL3}.  This means that for a significant interval of time the scale factor remains in the $[0,b]$ range, then rapidly traverses the classically forbidden $[b,d]$ interval via tunneling emerging at $a=d$ with $\dot{a}=0$.  Thereafter the scale factor grows slower according to the Friedmann equations (\ref{energy_integral0}) and (\ref{potential_energy0}).  This is an example of decay of the false vacuum \cite{Coleman}.  

The energy spectrum of quasi-stationary states consists of a collection of broadened levels (Figure \ref{pe}) of widths small compared to the separation between the levels.  The states are determined by solutions to the Schr\"odinger equation (\ref{SE1}) representing an outgoing spherical wave at infinity.  This complex boundary condition implies complex energy eigenvalues \cite{LL3},
\begin{equation}
\label{complex}
E\rightarrow E-\frac{i\Gamma}{2}
\end{equation}
where $\Gamma$ is the level width.  As a result, the time factor in the wave function (\ref{wave_function}) transforms according to
\begin{equation}
\label{time_factor}
\exp(-iEt)\rightarrow \exp(-iEt) \exp\left (-\frac{\Gamma t}{2}\right ).
\end{equation}
Then the probability to find the scale factor inside the $[0,b]$ range decreases with time as $\exp(-\Gamma t)$,  implying that $1/\Gamma$ is the lifetime of the quasi-stationary state \cite{LL3}. 

With the spectrum becoming complex (\ref{complex}), the pre-exponential part of the asymptotic outgoing wave function (\ref{asymptotic_solutions}) transforms according to
\begin{equation}
\label{a_factor}
a^{iE/\Omega-3/2}\rightarrow a^{iE/\Omega+\Gamma/2\Omega-3/2}.
\end{equation}
Corresponding wave function (\ref{asymptotic_solutions}) cannot be normalized, a property required to complement the exponentially decaying probability  \cite{LL3}.

Since for $E\ll E_{0}$ the barrier in Figure \ref{pe} is wide, the eigenvalues (\ref{complex}) can be determined by applying the semiclassical approximation \cite{LL3} to the one-dimensional Schr\"odinger equation (\ref{SE2}).  The starting point is the semiclassical wave function \cite{Migdal} 
\begin{equation}
\label{semiclassical_wave_function}
\chi(a<b)\sim \frac{1}{\sqrt{p(a)}}\sin\left (\int_{0}^{a}p(a')da'-\frac{\pi}{4}\right )
\end{equation} 
to be analytically continued into the $b<a<d$ region and then into the $a>d$ range.  This can be done with the help of the connection formulas summarized in Ref.\cite{Goldman}.  Requiring that there only is an outgoing wave for $a$ large, leads to the condition
\begin{equation}
\label{condition}
4i\tan\left (\int_{0}^{b}p(a)da\right )=D,
\end{equation}
\begin{equation}
\label{transmission_coefficient}
D(E)=\exp\left (-2\int_{b}^{d}|p(a)|da\right )
\end{equation}
where $D$ is the semiclassical transmission coefficient \cite{LL3}.  If the transmission is neglected, $D=0$ ($\Omega=0$), then Eq.(\ref{condition}) reduces to the quantization rule
\begin{equation}
\label{quantization_rule}
\int_{0}^{b}p(a)da=\pi (n_{r}+1)
\end{equation}
which recovers the Bohr spectrum (\ref{Bohr}) \cite{Migdal}.  Treating the right-hand side of Eq.(\ref{condition}) as a perturbation, correction to the energy eigenvalue of the form of Eq.(\ref{complex}) can be found with the following result for the level width \cite{comment}
\begin{equation}
\label{width}
\Gamma=\frac{1}{T(E)}D(E)
\end{equation}
where $T(E)$ is the classical period of motion,
\begin{equation}
\label{classical_period}
T(E)=\sqrt{2}\int_{0}^{b}\frac{da}{\sqrt{E-U(a)}},
\end{equation}
with the energy $E$ given by the Bohr spectrum (\ref{Bohr}).  The Gamow-type formula, Eq.(\ref{width}), has the well-known interpretation: transmission probability per unit time, $\Gamma$, is equal to the number of collisions per unit time $1/T$ against the barrier times the quantum-mechanical transmission probability $D$.   

Barrier penetration is known to be a consequence of energy fluctuations constrained by the energy-time uncertainty relation \cite{Cohen}.  This insight makes it possible to write down an expression for the time spent in the barrier region, also called the semiclassical time \cite{Cohen,Hagmann},
\begin{equation}
\label{transit_time}
\tau(E)=\int_{b}^{d}\frac{da}{\sqrt{2[U(a)-E]}}.
\end{equation}
For the potential energy (\ref{potential_energy1}) the quantities appearing in Eq.(\ref{classical_period}) can be computed in the $\Omega\ll 1$ limit with the following result for the level width (\ref{width})
\begin{equation}
\label{width_final}
\Gamma=\frac{2}{\pi}|E|^{3}\exp\left (-\frac{\pi |E|}{\Omega}\right )
\end{equation}
and the semiclassical time
\begin{equation}
\label{transit_time_final}
\tau=\frac{\pi}{2\Omega}.
\end{equation}
Comparing Eqs.(\ref{width_final}) and (\ref{transit_time_final}) we see that thanks to the fact that $\Omega\ll1$, the lifetimes $1/\Gamma$ of metastable states satisfying the condition of applicability of the semiclassical approximation, $D\ll1$ or $n_{r}\ll \Omega^{-1/2}$, are significantly larger than the tunneling times $\tau$.  During the tunneling time the scale factor increases by a large factor of $d/b=(2|E|^{3})^{1/2}/\Omega\gg1$.  Typical velocity of traversing the tunneling region estimated as $d/\tau\simeq \sqrt{|E|}\simeq 1/(n_{r}+1)$ is independent of $\Omega$.  

\section{Conclusions}

To summarize, we have demonstrated that quantum theory of Newtonian cosmology, despite its limitations, deals with the issue of the initial density singularity and naturally predicts evolution resembling the inflation scenario of modern cosmology.  The driving force of the evolution is the cosmological constant.  We hope that our analysis will motivate future studies that address critique of inflation as well as promote other avenues of inquiry.


\begin{thebibliography}{28}

\bibitem{Hubble}  E. Hubble, A relation between distance and radial velocity among extra-galactic nebulae?, Proc. Natl. Acad. Sci. U.S.A. \textbf{15}, 168 (1929).

\bibitem{Friedmann}  A. Friedman, \"Uber die Kr\"ummung des Raumes, Z. Phys. \textbf{10}, 377 (1922) [On the curvature of space, Gen. Relativ. Gravit. \textbf{31}, 1991 (1999)]; \"Uber die M\"oglichkeit einer Welt mit konstanter negativer Kr\"ummung des Raumes, Z. Phys. \textbf{21}, 326 (1924) [On the possibility of a world with constant negative curvature of space, Gen. Relativ. Gravit. \textbf{31}, 2001 (1999)].

\bibitem{Lem}G. Lema\^{\i}tre, Un univers homog\`ene de masse constante et de rayon croissant, rendant compte de la vitesse radiale des n\'ebuleuses extragalactiques?, Ann. Soc. Sci. Bruxelles A, \textbf{47}, 49 (1927); [A homogeneous universe of constant mass and increasing radius accounting for the radial velocity of extra-galactic nebulae?, Mon. Not. R. Astron. Soc. \textbf{91}, 483 (1931)].

\bibitem{Penrose}  S. W. Hawking and R. Penrose, \textit{The Nature of Space and Time}, Princeton University Press, Princeton, NJ, USA, 1996.

\bibitem{Kefir}  C. Kiefer, Conceptual Problems in Quantum Gravity and Quantum Cosmology, ISRN Math. Phys. 509316 (2013), and references therein;  \textit{Quantum Gravity}, (Oxford University Press, Oxford, UK, 3rd edition, 2012).

\bibitem{Harrison}  E. Harrison, \textit{Cosmology: The Science of the Universe}, 2nd Edition, (Cambridge University Press, 2022) Chapters 22 and 23.

\bibitem{Mukhanov} V. Mukhanov, \textit{Physical Foundations of Cosmology} (Cambridge University Press, 2005), Chapters 1 and 5.

\bibitem{Turok}  N. Turok, A critical review of inflation, Class. Quantum Grav. \textbf{19}, 3449 (2002).

\bibitem{opposition}  A. Ijjas, P. J. Steinhardt, and A. Loeb, Inflationary schism, Phys. Lett. B, \textbf{736} 142 (2014).

\bibitem{Coleman}  S. Coleman, Fate of the false vacuum: Semiclassical theory, Phys. Rev. D \textbf{15}, 2929 (1977).

\bibitem{LL2}   L. D. Landau and E. M. Lifshitz, \textit{The Classical Theory of Fields}, 4th ed., Course of Theoretical Physics Vol. II, (Pergamon. 1980), Sections 37, 87 and 99.

\bibitem{Milne}  E.A.Milne, A Newtonian Expanding Universe, Quart. J. Math. Oxford \textbf{5}, 64 (1934).
 
\bibitem{MM}  W. H. McCrea and E. A. Milne, Newtonian Universes and the Curvature of Space, Quart. J. Math. Oxford \textbf{5}, 73 (1934).	

\bibitem{McCrea0}  W. H. McCrea,  Relativity theory and the creation of matter, Proc. Roy. Soc. A \textbf{206}, 562 (1951). 

\bibitem{McCrea1}  W. H. McCrea, Newtonian Cosmology, Nature \textbf{175}, 466 (1955).

\bibitem{Zamora}  J. M. Romero and Zamora, Note on Quantum Newtonian Cosmology, https://arxiv.org/abs/gr-qc/0504072v1.  

\bibitem{Vieira15}  H. S. Vieira, and V. B. Bezerra, Quantum Newtonian cosmology and the biconfluent Heun functions, J. Math. Phys. \textbf{56}, 092501 (2015); https://doi.org/10.1063/1.4930871.

\bibitem{Vieira19}  H. S. Vieira, V. B. Bezerra, C. R. Muniz, and M.S Cunha, Some exact results on quantum Newtonian cosmology, J. Math. Phys. \textbf{60}, 102301 (2019); https://doi.org/10.1063/1.5086370.

\bibitem{Gouba}  L. Gouba, Quantum Newtonian cosmology revisited, https://arxiv.org/abs/2104.05524.

\bibitem{EBK}  E. B. Kolomeisky, Natural analog to cosmology in basic condensed matter physics, Phys. Rev. B \textbf{100}, 140301(R) (2019).

\bibitem{Bonnor}  W. B. Bonnor, Jean's formula for gravitational instability, MNRAS \textbf{117}, 104 (1957).

\bibitem{ZN}  Ya. B. Zeldovich and I. D. Novikov, \textit{Relativistic Astrophysics, 2: The Structure and Evolution of the Universe}, (University of Chicago Press, 1983).

\bibitem{LL3}  L. D. Landau and E. M. Lifshitz, \textit{Quantum Mechanics: Non-Relativistic Theory}, 3rd ed., Course of Theoretical Physics Vol. III, (Butterworth-Heinemann, Oxford, 1991), Section 18, Chapters V-VII and Section 134.

\bibitem{CMB}   Planck Collaboration, Planck 2018 results. VI. Cosmological parameters,  Astronomy and Astrophysics, \textbf{641}, A6 (2020);  \textit{Observable universe}, Wikipedia, The Free Encyclopedia, 13 May 2023.

\bibitem{Migdal}  A. B. Migdal, \textit{Qualitative Methods In Quantum Theory},  (CRC Press; 1st edition, 2019), Chapter 3.

\bibitem{Goldman}  I. I. Gol'dman and V. D. Krivchenkov, \textit{Problems in Quantum Mechanics}, (Dover Publications, 2010), Appendix  1.

\bibitem{comment}  There also are corrections to the Bohr spectrum (\ref{Bohr}) of order $\Omega^{2}$ that can be deduced from Eq.(\ref{quantization_rule}).  They are of no significance for our purpose and thus not discussed.   

\bibitem{Cohen}  B. L. Cohen, A Simple Treatment of Potential Barrier Penetration, American Journal of Physics \textbf{33}, 97 (1965).

\bibitem{Hagmann}  M. J. Hagmann, Transit Time for Quantum Tunneling, Solid State Communications, \textbf{82}, 867 (1992). 

\end{thebibliography}
\end{document}